\documentclass[sigconf]{acmart}

\AtBeginDocument{%
  \providecommand\BibTeX{{%
    \normalfont B\kern-0.5em{\scshape i\kern-0.25em b}\kern-0.8em\TeX}}}

\copyrightyear{2022}
\acmYear{2022}
\setcopyright{acmcopyright}\acmConference[KDD '22]{Proceedings of the 28th ACM SIGKDD Conference on Knowledge Discovery and Data Mining}{August 14--18, 2022}{Washington, DC, USA}
\acmBooktitle{Proceedings of the 28th ACM SIGKDD Conference on Knowledge Discovery and Data Mining (KDD '22), August 14--18, 2022, Washington, DC, USA}
\acmPrice{15.00}
\acmDOI{10.1145/3534678.3539123}
\acmISBN{978-1-4503-9385-0/22/08}

\usepackage{multirow}
\usepackage{footmisc}
\usepackage{subcaption,siunitx,booktabs}
\usepackage{caption}
\usepackage{bbm}
\usepackage{xcolor}
\usepackage[ruled]{algorithm2e}
\usepackage{balance}
\usepackage{graphicx}
\usepackage{enumitem}
\usepackage{todonotes}
\usepackage{tablefootnote}

\renewcommand{\todo}[1]{\iffalse #1 \fi[TODO]}

\SetKwInput{KwInput}{Input}                
\SetKwInput{KwOutput}{Output}              

\begin{document}

\title{Intelligent Request Strategy Design in Recommender System}

\author[X. Qian, Y. Xu, F. Lv, S. Zhang, Z. Jiang, Q. Liu, X. Zeng, T. Chua, F. Wu]{
    Xufeng Qian$^{1}$, Yue Xu$^{1}$, Fuyu Lv$^{1}$, Shengyu Zhang$^{2*}$, Ziwen Jiang$^{1}$, Qingwen Liu$^{1}$, Xiaoyi Zeng$^{1}$, Tat-Seng Chua$^{5}$, Fei Wu$^{3,4}$
}

\affiliation{
	$^1$ Alibaba Group, Hangzhou, China \ $^2$ Institute of Artificial Intelligence, Zhejiang University, Hangzhou China \country{}
}
\affiliation{
     $^3$ Shanghai Institute for Advanced Study of Zhejiang University, Shanghai, China \country{}
}
\affiliation{
	$^4$ Shanghai AI Laboratory, Shanghai, China \ $^5$ National University of Singapore, Singapore \country{}
}

\email{
	{xufeng.qxf, guyue.yuexu, fuyu.lfy, xiangsheng.lqw}@alibaba-inc.com
}
\email{zjujzw0303@gmail.com, yuanhan@taobao.com,{sy_zhang, wufei}@zju.edu.cn, dcscts@nus.edu.sg}
  

\thanks{* Shengyu Zhang is the corresponding author. }

\renewcommand{\shortauthors}{Xufeng Qian et al.}

\renewcommand{\authors}{Xufeng Qian, Yue Xu, Fuyu Lv, Shengyu Zhang, Ziwen Jiang, Qingwen Liu, Xiaoyi Zeng, Tat-Seng Chua, Fei Wu}

\newcommand{\etal}{\textit{et al}.}
\newcommand{\ect}{\textit{ect}}
\newcommand{\ie}{\textit{i}.\textit{e}.}
\newcommand{\eg}{\textit{e}.\textit{g}.}
\newcommand{\wrt}{\textit{w}.\textit{r}.\textit{t}. }
\newcommand{\vpara}[1]{\vspace{0.05in}\noindent\textbf{#1 }}
\newcommand\norm[1]{\left\lVert#1\right\rVert}

\begin{abstract}

Waterfall Recommender System (RS), a popular form of RS in mobile applications, is a stream of recommended items consisting of successive pages that can be browsed by scrolling. 
In waterfall RS, when a user finishes browsing a page, the edge (\eg, mobile phones) would send a request to the cloud server to get a new page of recommendations, known as the paging request mechanism.
RSs typically put a large number of items into one page to reduce excessive resource consumption from numerous paging requests, which, however, would diminish the RSs' ability to timely renew the recommendations according to users' real-time interest and lead to a poor user experience.
Intuitively, inserting additional requests inside pages to update the recommendations with a higher frequency can alleviate the problem.
However, previous attempts, including only non-adaptive strategies (\eg, insert requests uniformly), would eventually lead to resource overconsumption. 
To this end, we envision a new learning task of edge intelligence named Intelligent Request Strategy Design (\textbf{IRSD}). It aims to improve the effectiveness of waterfall RSs by determining the appropriate occasions of request insertion based on users' real-time intention. 
Moreover, we propose a new paradigm of adaptive request insertion strategy named Uplift-b\textbf{a}sed On-e\textbf{d}ge Sm\textbf{a}rt \textbf{Request} Framework (\textbf{AdaRequest}). AdaRequest 1) captures the dynamic change of users' intentions by matching their real-time behaviors with their historical interests based on attention-based neural networks. 2) estimates the counterfactual uplift of user purchase brought by an inserted request based on causal inference. 3) determines the final request insertion strategy by maximizing the utility function under online resource constraints. We conduct extensive experiments on both offline dataset and online A/B test to verify the effectiveness of AdaRequest. Remarkably, AdaRequest has been deployed on the Waterfall RS of Taobao and brought over \textbf{3\%} lift on Gross Merchandise Value (GMV).

\end{abstract}

\ccsdesc[500]{Information systems~Mobile information processing systems}

\keywords{Recommender Systems, Intelligent Request Strategy, User Intention}


\maketitle

\section{Introduction}

\begin{figure}[!t] \begin{center}
    \includegraphics[width=\columnwidth]{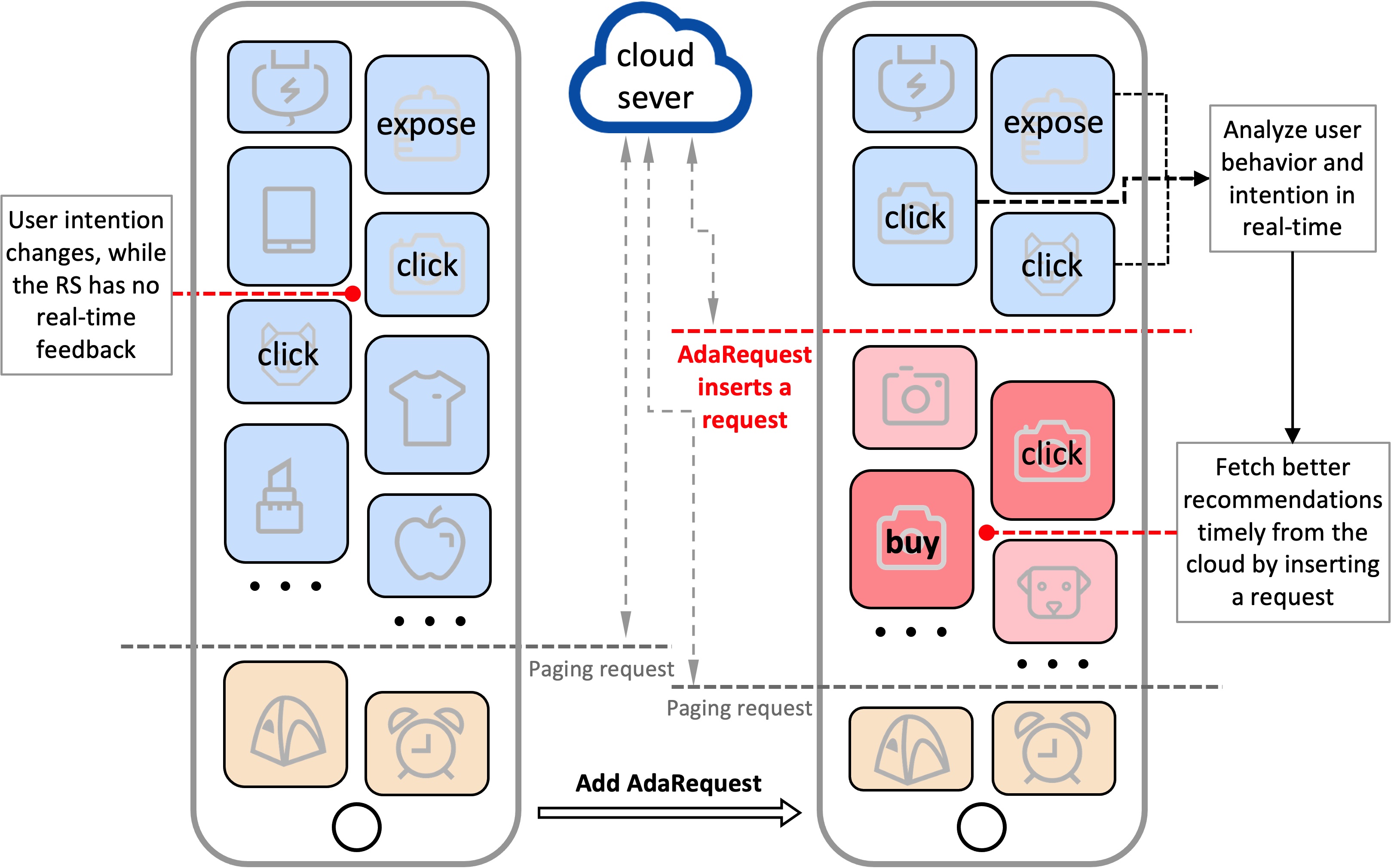}
    \caption{
    	An illustration of AdaRequest in waterfall RS with an example. \textit{AdaRequest inserts a request timely to fetch more cameras from the cloud after the user clicks on a camera, thus stimulating more purchases from the user.}
	}
\label{fig:pagerequest}
\vspace{-1.0em}
\end{center} 
\end{figure}

Modern Recommender Systems (RS)~\cite{Li_Lv_Jin_Lin_Yang_Zeng_Wu_Ma_2021,Gong_Jiang_Feng_Hu_Zhao_Liu_Ou_2020,Yao_Wang_Jia_Han_Zhou_Yang_2021,DBLP:conf/sigir/ZhangYZC021,DBLP:conf/www/ZhangYYLFZC022} often expose a list of items to the users on the edges (\eg, mobile phones) where the items are sorted by the relevance to users' interests.
\textbf{Waterfall RS}, a popular form of RS, is a stream of recommended items consisting of successive pages that can be browsed by scrolling, as shown in \textit{Fig.}~\ref{fig:pagerequest},
\eg, the homepage commodity RS \textit{Guess You Like} in Mobile Taobao~\cite{Gong_Jiang_Feng_Hu_Zhao_Liu_Ou_2020} and the homepage video RS in Mobile YouTube~\cite{Covington_Adams_Sargin_2016}. 
In traditional waterfall RSs, the number of items in each page (\textit{page size}) is generally fixed.
Once a user finishes browsing a page and continuously scrolls down, the edge would communicate with the cloud server to get a new page of recommendations by initiating a \textbf{request}, which is called \textbf{paging request}, as shown in \textit{Fig.}~\ref{fig:pagerequest}. 
In paging request, the cloud would re-analyze the user's new intention, generate the next page of recommendations, and then send the page back to the edge for subsequent exposure.
In order to reduce the vast computational resource consumption arising from numerous paging requests, RSs typically set the page size to a large one (\eg, 50 items).
However, the paging request mechanism with a large page size seriously undermines RSs' ability to timely update the recommendations according to users' real-time interest.
For instance, if a user suddenly changes his intention when he/she reaches the middle of a page 
and clicks on an item that the user has never followed before, 
the RS with a non-adaptive paging request mechanism would not provide any real-time feedback.
Therefore, the user may get bored and quit early before he/she browsed through the whole page to get the updated recommendations.
Empirically, we provide statistic evidence from the RS \textit{Guess You Like}. 
As shown in \textit{Fig.}, there is a significant decrease in users' click-through-rate (CTR) when they are browsing items before the next paging request, which reflects their loss of interest, and a sharp increase of CTR appears after the recommendation is updated via the paging request. 
The RS with a non-adaptive paging request mechanism cannot provide real-time feedback on the decline of user CTR and analyze user intention timely to renew recommendations, resulting in a lower average CTR and a poorer user experience.

Intuitively, inserting additional requests between adjacent paging requests to update the recommendations with a higher frequency can alleviate the above problem.
However, previous attempts in this direction include only simple non-adaptive strategies, \eg, uniformly or randomly inserting requests between paging requests. 
Take the example of randomly inserting requests with a certain probability in the RS \textit{Guess You Like}. As shown in \textit{Fig.}, when additional requests are increasingly imposed with a higher probability, there is a consistent improvement in the number of transactions, indicating that inserting more additional requests is indeed effective.
However, non-adaptive request strategies pose another problem: 
massive additional requests from an enormous number of users would consume vast amounts of bandwidth and computational resources.
Since non-adaptive request insertion strategies do not take users' intention into consideration, many users do not purchase or browse more after a request is imposed, resulting in a significant waste of request resources.


To solve the problem above, we propose a brand new task of edge intelligence: Intelligent Request Strategy Design (\textbf{IRSD}).
It requires us to design an adaptive request insertion strategy to determine the appropriate occasions of request insertion to trade off the \textit{uplift}~\cite{lo2002true,hansotia2002incremental} in positive user feedback (\eg, purchase), aka the \textbf{request reward}, and the computational resource consumption.
To the best of our knowledge, this is the first attempt to investigate the task \textbf{IRSD} in the literature of waterfall RS. The major challenges of this task are summarized as the following:
\begin{itemize}[leftmargin=*]
	\item Capturing the dynamic change of the user's intention is non-trivial since the user's intention changes implicitly~\cite{Guo_Hua_Jia_Fang_Zhao_Cui_2020,guo2019buying}.
	\item  Since the feedback of individual users with and without requests cannot be obtained simultaneously, the groundtruth \textit{uplift} of positive user feedback brought by requests cannot be observed counterfactually, making the \textit{uplift} estimation a big challenge.
	\item Trading off the sum of all the request rewards and the resource consumption is challenging, especially when the number of users varies dynamically over time.
\end{itemize}

\begin{figure}[!t] \begin{center}
\begin{subfigure}{.235\textwidth}
	\includegraphics[width=0.95\linewidth]{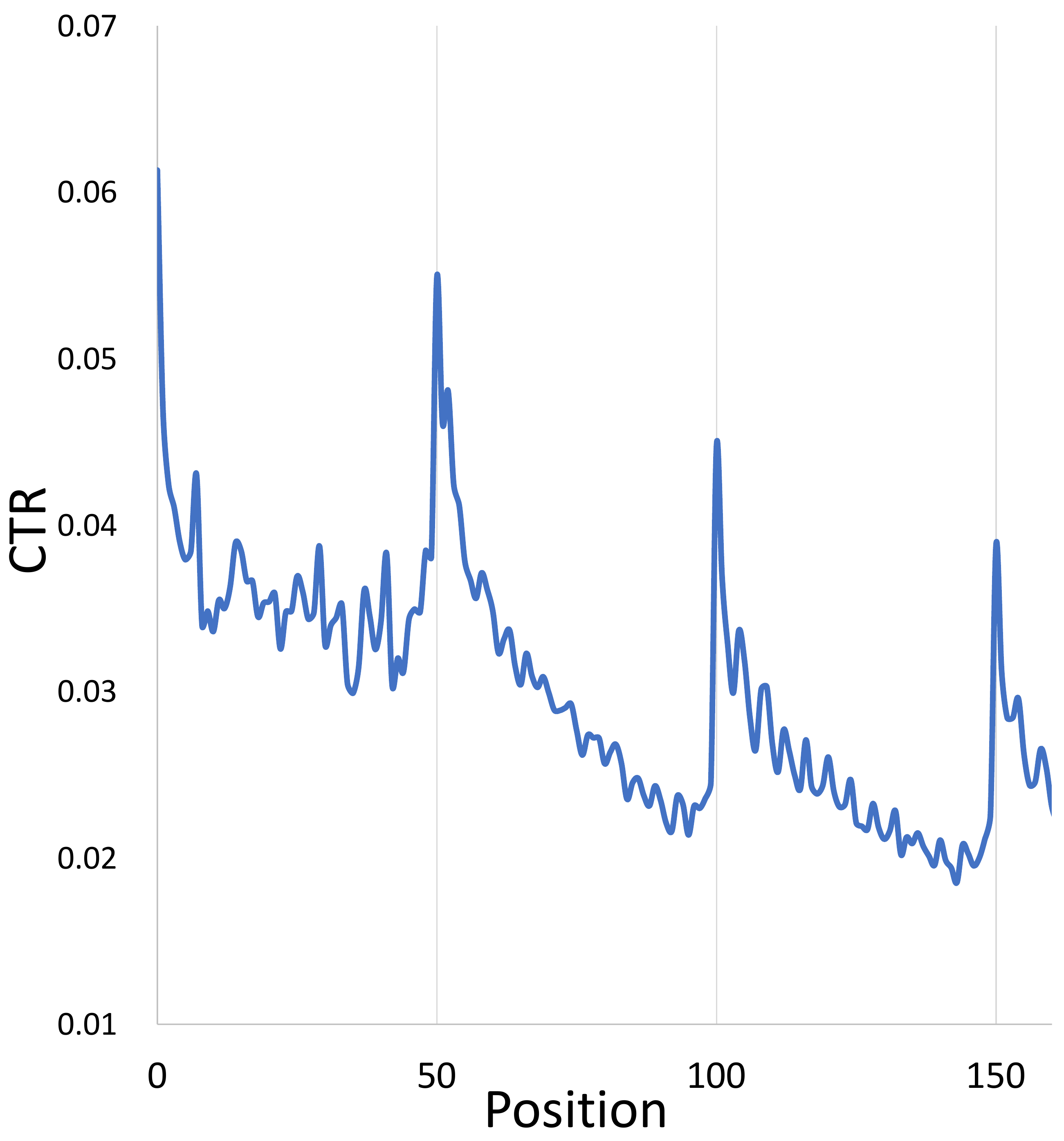}
\label{fig:ctrrequestleft}
\end{subfigure}
\begin{subfigure}{.235\textwidth}
	\includegraphics[width=0.95\linewidth]{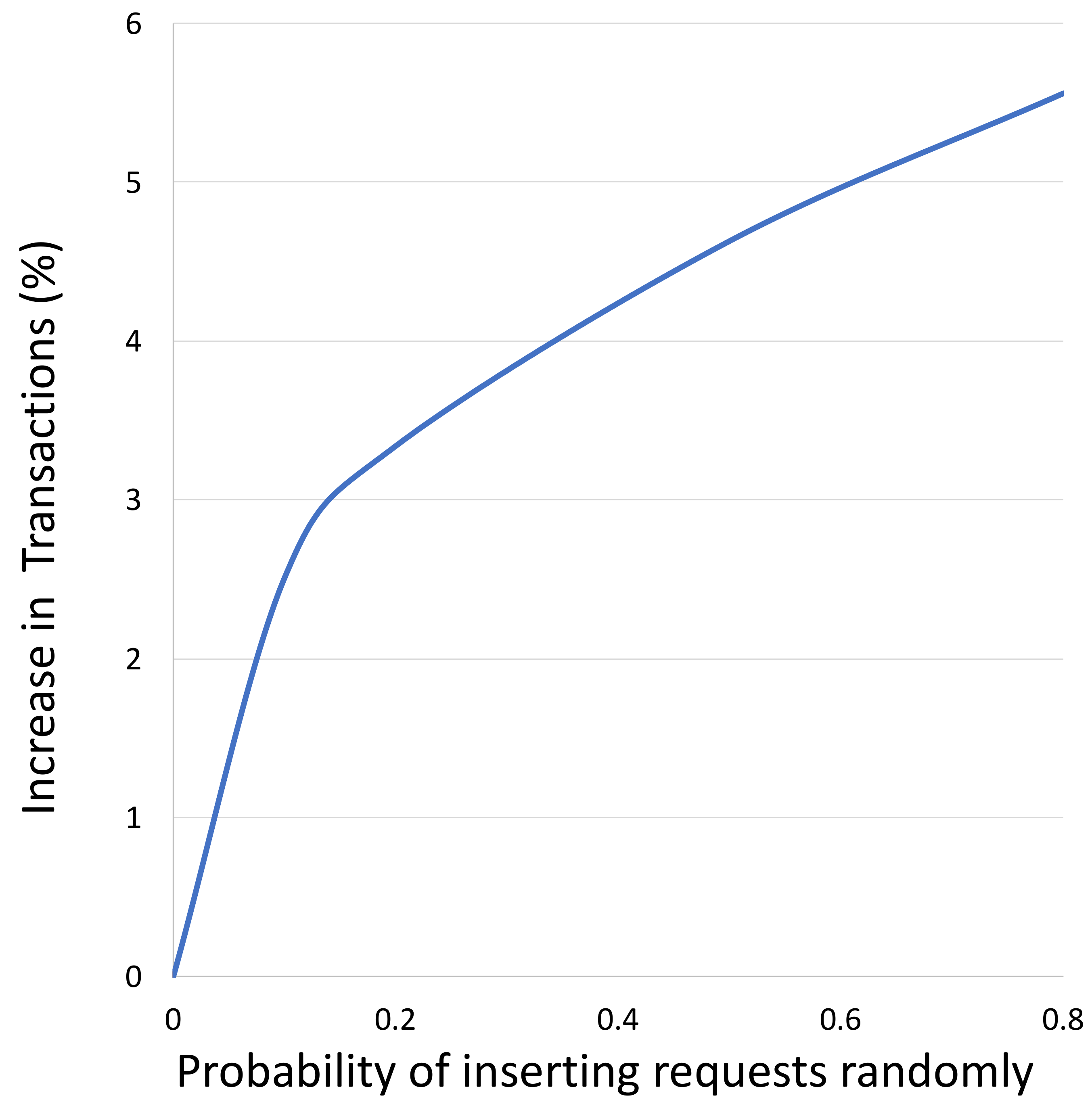}
\label{fig:ctrrequestright}
\end{subfigure}
    \caption{
    	(Left) How Click-through-rate (CTR) correlates with paging requests. There is a paging request per 50 items. (Right) The increase in the transactions when inserting an increasing number of requests between two paging requests.
    	}
    \label{fig:ctrrequest}
\vspace{-0.4cm}
\end{center} 
\end{figure}


To address the above challenges, we propose a new paradigm of adaptive request insertion strategy named the Uplift-b\textbf{a}sed On-e\textbf{d}ge Sm\textbf{a}rt \textbf{Request} Framework (\textbf{AdaRequest}).
AdaRequest is deployed on \textbf{edge}, enabling it to collect and analyze user behavior in an ultra-real-time manner.
AdaRequest consists of three critical components, \ie, comprehensive user behavior modeling (\textbf{CUBE}), counterfactual request reward estimator (\textbf{CREST}), and dynamic request programming (\textbf{DRP}). 
Specifically, \textbf{CUBE} performs user real-time intention mining by comprehensively modeling users' diverse real-time behavior sequences, long-term behavior sequences, and interaction commodity sequences.
Since the candidate items for subsequent exposure on edge are selected by the cloud based on user's historical interest, by comparing user's real-time intention with the historical interest, \textbf{CUBE} evaluates whether the user's intention has changed.
For the lack of the groundtruth uplift of positive user feedback, we divide the population into the TREATMENT group imposing additional requests and the CONTROL group without additional requests. \textbf{CREST} aims to model the user behavior of the two groups separately and estimate the uplift by causal inference.
To avoid the overconsumption of resources, typically, a budget would be set for the resources of additional requests.
\textbf{DRP} would dynamically select a fraction of cases with a higher ratio of the request reward to the resource consumption for requests insertion, maximizing the sum of all the request rewards under limited resources.
Remarkably, 
AdaRequest has been deployed on the \textit{Guess You Like} in Mobile Taobao and brought over \textbf{3\%} lift on GMV, compared to the previous online baseline through A/B test. 
Specifically, the previous online baseline is a request strategy that uniformly inserts a constant number of requests between paging requests. 
To summarize, this paper makes the following contributions:
\begin{itemize}[leftmargin=*]
	\item We propose and investigate a new task \textbf{IRSD} in the literature of waterfall RS. 
	\textbf{IRSD} will play a critical role in scenarios that require trading off the request rewards and resource consumption, and can be widely studied in most RSs and search systems.
	\item We devise a novel paradigm \textbf{AdaRequest} to accomplish the \textbf{IRSD} task, 
	where \textbf{CUBE} captures the dynamic change of users' intentions, \textbf{CREST} estimates the counterfactual uplift of positive user feedback by causal inference, and \textbf{DRP} maximizes the overall rewards of requests dynamically under limited resources.
	\item 
	We conduct extensive offline and online experiments on a real-world RS, where the qualitative and quantitative results jointly demonstrate the effectiveness of \textbf{AdaRequest}. 
	Remarkably, AdaRequest deployed on Mobile Taobao improved the GMV by over \textbf{3\%} percent compared to the previous online baseline.
\end{itemize}

\section{Related Works}

\vpara{Edge Computing.} 
Edge computing has merits in reducing latency~\cite{Faruque_Vatanparvar_2016}, personalizing services \cite{He_Wei_Chen_Tang_Zhou_Zhang_2018,Iqbal_Butt_Shafique_Talib_Umer_2018,Lin_Yang_2018}, resource optimization \cite{Jia_Hu_Zeng_Xu_Yang_2018,Li2022End2End}, and strengthening privacy and security~\cite{Lyu_Nandakumar_Rubinstein_Jin_Bedo_Palaniswami_2018,Sohal_Sandhu_Sood_Chang_2018,Wang_Zhou_Chen_Wang_Liu_Liu_2018}. Edge computing for recommendation is still a nascent research area \cite{Gong_Jiang_Feng_Hu_Zhao_Liu_Ou_2020,Guo_Hua_Jia_Fang_Zhao_Cui_2020,Yao_Wang_Jia_Han_Zhou_Yang_2021,Yao_Zhang_Yao_Wang_Ma_Zhang_Chu_Ji_Jia_Shen_2021,DBLP:journals/corr/abs-2109-12314}. \cite{Guo_Hua_Jia_Fang_Zhao_Cui_2020} monitor users' timely multi-intentions on edge, such as whether a particular user will buy some goods within an hour, via binary classification. 
\cite{Yao_Wang_Jia_Han_Zhou_Yang_2021} introduce a collaborative learning framework to make on-edge training and on-cloud training benefit from each other. 
\cite{Gong_Jiang_Feng_Hu_Zhao_Liu_Ou_2020} 
focus on item re-ranking in the waterfall RS, providing more accurate recommendation while reducing the communication overhead on-edge. Different from these works, we aim to devise intelligent request strategies that actively request the server to update the candidate item pool on edge. In technique, we differ existing works by comprehensively understanding users' multi-grained behavior features and estimating the uplift rather than the absolute revenue for decision-making.

\vpara{Uplift Modeling.} Uplift modeling aims to estimate the incremental effect of taking action on an outcome through causal inference~\cite{wu2022learning,DBLP:conf/mm/ZhangJWKZZYYW20,li2022deconfounded}, which has broad applications varying from marketing~\cite{guelman2012random} to medical~\cite{alemi2009improved,li2020ib,zhang2022boostmis} domains. 
Class Transformation methods \cite{shaar2016pessimistic,athey2015machine}directly estimate the uplift based on two assumptions, \ie, binary outcome variable and balanced dataset
between control and treatments, which might not be satisfied in real-world scenarios.  
Two-Model \cite{athey2015machine,zaniewicz2013support,nie2021quasi} is a straightforward solution that models the treatment group data and control group data separately. Direct estimation modifies existing machine learning algorithms for uplift prediction, such as logistic regression~\cite{lo2002true}, SVM~\cite{zaniewicz2013support}, and tree-based approaches~\cite{radcliffe2011real,rzepakowski2012decision}. In this paper, we mainly follow the Two-Model line of research but enable backbone parameter-sharing, which leads to more robust and consistent user representation learning. Another advantage of our model is the uplift prediction head, which prevents the twice inference in Two-Model methods.

\vpara{Occasion-based Recommendation.} Recommendation occasion refers to the time point where users' preferences have significantly changed. Occassion modeling~\cite{DBLP:conf/wsdm/WangLHCCH20,DBLP:conf/www/WangLF0LC22} is tightly related to sequential recommendation~\cite{DBLP:conf/mm/XunZZZZLH0C021,DBLP:conf/www/LuZHWYZW21}. Typically, OAR~\cite{DBLP:conf/wsdm/WangLHCCH20} leverage the attention mechanisms~\cite{DBLP:conf/nips/VaswaniSPUJGKP17,zhang2019frame} elicit global occasion signal (\eg, festival) and local occasion signal (\eg, birthday), and accordingly make recommendations. 
Different from previous works, we model personalized occassion on edge, pursuing real-time analysis. 

\section{Methods}


We first introduce the notations used in this paper and formulate the problem of the task \textbf{IRSD}.
The algorithm for \textbf{IRSD} would be deployed on edge to determine whether to insert a request or not whenever it is invoked.
Let $\mathcal{X} = \{ X_{1},\dots,X_{N} \}$ denote a collection of $N$ cases where the algorithm is invoked and $\mathcal{C^{X}} = \{ c^{X}_{1},\dots,c^{X}_{N} \}$ denote the corresponding local contextual information in the waterfall RS. Let $\mathcal{U^{X}} = \{ u^{X}_{1},\dots,u^{X}_{N} \}$ denote the users who invoke the algorithm and $\mathcal{B}^{u} = \{ b^{u}_{1},\dots,b^{u}_{N} \}$ denote the users' behavior history. To avoid the overconsumption of computational resources, we typically set a upper bound $\theta$ for the resource consumption of additional requests over a period of time. 
Given all the cases $\mathcal{X}$, the local contextual information $\mathcal{C^{X}}$, the corresponding users $\mathcal{U^{X}}$ and their historical behaviors $\mathcal{B}^{u}$, we aim to find the best algorithm $\mathcal{F}$ to maximize the overall positive feedback from all the users while keeping the sum of the resource consumption brought by additional requests within the upper bound $\theta$, i.e.,
\begin{gather}
\underset{\mathcal{F}}{\text{Maximize}} \sum_{i \in [1, N]} \mathcal{S}(\mathcal{F}(c^{X}_{i}, u^{X}_{i}, b^{u}_{i}), X_{i}), \\
\text{subject to} \sum_{i \in [1, N]} \mathcal{R}(\mathcal{F}(c^{X}_{i}, u^{X}_{i}, b^{u}_{i}), X_{i}) < \theta.
\end{gather}
where $\mathcal{F}$ denotes the algorithm of which the output is a binary request decision, $\mathcal{S}$ denotes the positive user feedback, $\mathcal{R}$ denotes the additional resource consumption. 
We conducted experiments in the commodity RS \textit{Guess You Like} in Mobile Taobao and used the occurrence of purchase behaviors as the positive user feedback. 

\subsection{Overview of AdaRequest} \label{sec:overview}

\begin{figure*}[!t] 
\begin{center}
	\includegraphics[width=0.98\linewidth]{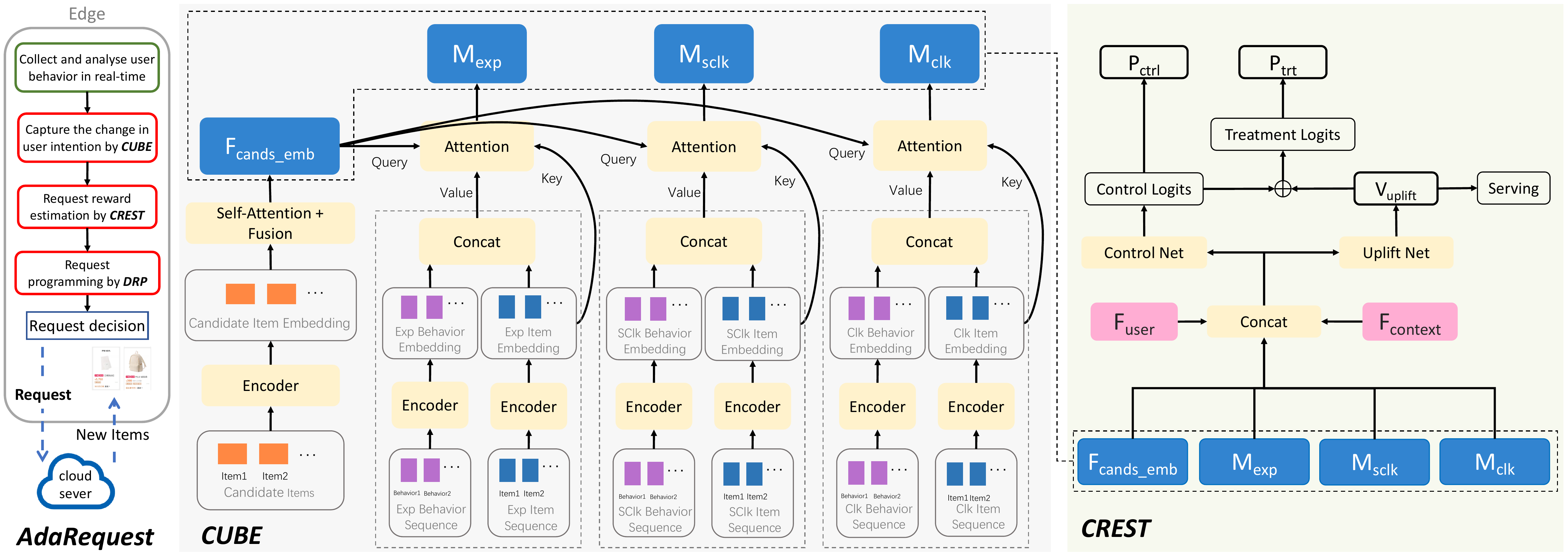}
    \caption{ Overview of AdaRequest Framework.
    	}
    \label{fig:CUBE_CURVE}
\vspace{-1.5em}
\end{center} 
\end{figure*}

To solve the above problem, we propose the AdaRequest framework.
The AdaRequest deployed on edge can collect and store user behavior in real-time.
As shown in the \textit{Fig.} \ref{fig:CUBE_CURVE}, first, the module \textbf{CUBE} in \textit{Sec} \ref{sec:CUBE} would perform user real-time intention mining by comprehensively analyzing and modeling the users' behavior sequences and the interaction commodity sequences. 
Candidate items, stored on edge for subsequent exposure as recommendations, are selected by the cloud server based on the user's historical interest.
\textbf{CUBE} would infer the user's historical interest based on the candidate items and compare it with the user's real-time intention by \textit{multi-head\ attention} to capture the dynamic change in the user's intention.
Then, since the additional requests are expected to bring in more online purchases, the module \textbf{CREST} in \textit{Sec} \ref{sec:CREST} would estimate the \textit{uplift} of purchase brought by each request, \ie, the request reward, by taking the contextual information and the output of \textbf{CUBE} as input. 
Since the groundtruth of the purchase uplift cannot be observed, \textbf{CREST} would model the user behavior in the CONTROL and TREATMENT groups by two separate branches and perform the \textit{uplift} estimation by causal inference.
After that, the \textbf{DRP} algorithm in \textit{Sec} \ref{sec:DRP} would make the final decision of request insertion by dynamically selecting a fraction of cases with a higher ratio of the request reward to the resource consumption, aiming to maximize the sum of all the request rewards under limited resources.
Finally, if AdaRequest decides to insert a request, the edge would communicate with the cloud and upload the user's real-time behaviors. The cloud would then re-analyze the user's intention and send back new items that better match the user's new interests for subsequent recommendations.


\vpara{Feature System.} 
We refer to a $\mathrm{session}$ as the period from the time a user enters the waterfall RS to the moment he/she leaves. 
For case $X_{i}$, we divide all the features into the following groups:
\begin{itemize}[leftmargin=*]
\item User attribute features $F_{user}$
\item Features of candidate items $F_{cands}$
\item User's fine-grained behaviors and interacted items
    \begin{itemize}
    \item User's browsing history in the current session $F_{exp}$
    \item User's click history in the current session $F_{sclk}$
    \item User's long-term click history $F_{clk}$
    \end{itemize}
\item Coarse-grained features of the current session's context $F_{context}$
\end{itemize}
where fine-grained user behavior features $b^{u}_{i} = \{F_{exp}, F_{sclk}, F_{clk}\}$, and coarse-grained context features $c^{X}_{i} = \{F_{user}, F_{cands}, F_{context}\}$. Notably, each of the feature groups $F_{exp}, F_{sclk}, F_{clk}$ includes two chronological sequences: a user behavior sequence and an interacted items sequence. 
In these feature groups, $F_{sclk}$ and $F_{clk}$ indicate users' real-time interest and long-term interest, respectively.
$F_{exp}$ is used to indicate the items boring the user after exposure since users' negative feedback is also valuable.
Features in $F_{context}$ indicate the user's satisfaction in the recommendations they have just viewed, \eg, the depth of view. More details on the feature can be found in Appendix \ref{sec:featdetail}.

\subsubsection{\textbf{Comprehensive User Behavior Modeling (CUBE)}} \label{sec:CUBE}

The module \textbf{CUBE} aims to infer whether users' intentions have changed from their real-time behaviors. Since the candidate items are selected by the cloud based on the user's interest history, \textbf{CUBE} would match the candidate items with the items that users have just interacted with and the corresponding user behaviors. If they are less matched, it indicates that the candidate items will not satisfy users' current intentions, which further implies that users' intentions may have changed, and thus an additional request at this moment may lead to a significant purchase uplift.

Specifically, as shown in the \textit{Fig.} \ref{fig:CUBE_CURVE}, \textbf{CUBE} takes $F_{exp}$, $F_{sclk}$, $F_{clk}$ and $F_{cands}$ as the input data and models the complex and diverse user behavior in an interpretable way. Different behavior sequences are heterogeneous both in structures and semantics. For example, $F_{sclk}$ and $F_{clk}$ reflect short-term and long-term user interests, respectively. Besides, $F_{exp}$ is far denser than $F_{clk}$, and might dominate the final representation if encoded jointly. Therefore, feature groups $F_{exp}, F_{sclk}, F_{clk}$ of different types are encoded separately with three structurally identical but parameter-independent model branches. 
Similarly, the user behavior sequence and the interacted item sequence in each of these three feature groups are encoded separately by two encoders in each model branch.
Taking $F_{sclk}$ as an example, 
for the discrete values in $F_{sclk}$, we first transform them into embeddings by retrieving the embedding matrices, while the continuous values in $F_{sclk}$ would be input directly. 
After that, as \textit{Eq.} \ref{eq:encoder0}-\ref{eq:encoder2}, we encode the user behavior sequence $F_{sclk\_beh}$ and the interacted item sequence $F_{sclk\_item}$ in $F_{sclk}$ into two embedding sequences $\widehat{F}_{sclk\_beh}$ and $\widehat{F}_{sclk\_item}$ with different Encoders, which two are then concatenated to obtain $F_{sclk\_embs}$. Since $F_{sclk\_beh}$ and $F_{sclk\_item}$ are two chronological sequences, we use Gate Recurrent Unit~\cite{cho2014learning} \textit{(GRU)} as the Encoder to encode the temporal evolution of user behavior.
\begin{gather}
        \widehat{F}_{sclk\_beh} = Encoder(F_{sclk\_beh}), \label{eq:encoder0} \\ 
        \widehat{F}_{sclk\_item} = Encoder(F_{sclk\_item}), \label{eq:encoder1} \\
        F_{sclk\_embs} = \widehat{F}_{sclk\_beh} \rVert \widehat{F}_{sclk\_item}, \label{eq:encoder2}
    \end{gather}
where $\rVert$ denotes the concatenation operation. Similarly, we can derive embedding sequences $\widehat{F}_{exp\_beh}$, \ $\widehat{F}_{exp\_item}$, \ $F_{exp\_embs}$ and $\widehat{F}_{clk\_beh}$, \ $\widehat{F}_{clk\_item}$, \ $F_{clk\_embs}$ from $F_{exp}$, $F_{clk}$ respectively. 
\begin{gather}
    F_{cands\_emb} = Fusion(\textit{self-attention}(Encoder(F_{cands}))). \label{eq:selfattn}
\end{gather}
After that, we encode the sequence of candidate items $F_{cands}$ to extract a single embedding $F_{cands\_emb}$ representing the overall characteristics of the candidate items, as \textit{Eq.} \ref{eq:selfattn}. 
Since candidate items $F_{cands}$ have been ranked by the cloud according to their relevance to the user's preferences, we use \textit{GRU} as the Encoder to encode the sequential relevance in $F_{cands}$.
Besides, \textit{mean-pooling} operation is used as the Fusion function. 

Finally, as the CUBE part in \textit{Fig.} \ref{fig:CUBE_CURVE}, we use \textit{multi-head\ attention} as the operator for matching candidate items with user behavior and interacted item sequence. $F_{cands\_emb}$ is used as the \textit{query} in \textit{multi-head\ attention} to match the interacted item embedding sequences to evaluate whether the candidate items will satisfy the user's real-time intention. Taking $F_{sclk}$ as an example, as in \textit{Eq.} \ref{eq:multiattn0}-\ref{eq:multiattn1}, 
\begin{gather}
    \alpha_{i} = \frac{\exp(\mathit{Match}(F_{cands\_emb}, \widehat{F}_{sclk\_item}^{i}))}{\sum_{j \in [0, L)}\limits \exp(Match(F_{cands\_emb}, \widehat{F}_{sclk\_item}^{j}))} \label{eq:multiattn0}, \\ 
    \mathbf{M}_{sclk} = \sum_{i \in [0, L)}\limits \alpha_{i} F_{sclk\_embs}^{i}, \label{eq:multiattn1}
\end{gather}
where the weight $\alpha_{i}$ denotes the matching score between the $i^{th}$ item in the interacted item sequence $F_{sclk\_item}$ and the candidate items $F_{cands}$. 
The resulting embedding $\mathbf{M}_{sclk}$ contains clues about whether the candidate items match the user's real-time intention.
$L$ is the length of sequences $\widehat{F}_{sclk\_item}$ and $F_{sclk\_embs}$. 
$\mathit{Match}(\cdot)$ refers to the matching score function, which measures the similarity between the recommended candidate items and the items meeting user’s real-time interests. This paper uses the inner product as the matching score function for simplicity.
Similar to the derivation of $\mathbf{M}_{sclk}$, we can obtain the embedding $\mathbf{M}_{exp}$ and $\mathbf{M}_{clk}$ which contains the information about whether the candidate items match the items boring the user and user's long-term interest, respectively.

\subsubsection{\textbf{Counterfactual Request Reward Estimator (CREST)}} \label{sec:CREST}

The module \textbf{CREST} aims to estimate the purchase uplift brought by inserting an additional request, \ie, the request reward.
However, since the feedback of individual users with and without requests cannot be obtained simultaneously at a particular time point, one cannot counterfactually observe the groundtruth of purchase uplift.
To bridge the gap, we propose to train two separate prediction networks, \ie, the Control Net, which predicts the purchase rate if a request is not triggered, and the Uplift Net, which estimates the request reward. The Control Net together with the Uplift Net predicts the purchase rate if a request is triggered, as depicted in Fig. \ref{fig:CUBE_CURVE}. 
To accommodate the training of two prediction networks, we divide the experimental users randomly into two groups in data collection: 1) the CONTROL group where no additional request is triggered and 2) the TREATMENT group where additional requests are triggered randomly. 
The Control Net is trained with the data from the CONTROL group, and the Uplift Net is trained with the data from the CONTROL group and TREATMENT group.
During inference, only the Uplift Net would be run to estimate the request reward, which serves as the basis for deciding whether a request should be triggered.

To summarize the modeling via the language of causality~\cite{gutierrez2017causal}, in the context satisfying the Conditional Independent Assumption $(CIA)$, \textbf{CREST} aims to estimate the Conditional Average Treatment Effect $(CATE)$ of the additional request, which can be written as:
\begin{gather}
CIA: \{Y_i(0), Y_i(1)\} \bot z_i \vert X_i, \label{eq:cate0} \\
CATE_P(X_i) = E[Y_i(1) \vert X_i]-E[Y_i(0) \vert X_i]. \label{eq:cate1}
\end{gather}
where $CATE_P(X_i)$ denotes the uplift of purchase stimulated by a request in the case $X_{i}$, $z_i$ denotes whether a request is applied, and $Y_i(1)$, $Y_i(0)$ denote whether a user makes purchase with or without an inserted request, respectively.

Specifically, as presented by the CREST part of \textit{Fig.} \ref{fig:CUBE_CURVE}, we first concatenate the outputs of \textbf{CUBE} with $F_{user}$, $F_{context}$ to obtain the fusion embedding $F_{fusion}$. 
Then, we take $F_{fusion}$ as input and build two networks named Control-net and Uplift-net.
Control-net models the user behavior from the CONTROL group and outputs the purchase logits $Logits_{ctrl}$ when there is no additional request. 
Uplift-net models the difference between the user behavior in the CONTROL group and the TREATMENT group, and outputs the logits of the purchase uplift $V_{uplift}$.
The above procedures can be represented as:
\vspace{-0.5em}
\begin{gather}
F_{fusion} = F_{cands\_emb} \rVert \mathbf{M}_{exp} \rVert \mathbf{M}_{sclk} \rVert \mathbf{M}_{clk} \rVert F_{user} \rVert F_{context}, \label{eq:curve0} \\
Logits_{control} = \text{Control-net}(F_{fusion}), \label{eq:curve1} \\
V_{uplift} = \text{Uplift-net}(F_{fusion}), \label{eq:curve2}
\end{gather}
We use two \textit{MLPs} as the Control-net and the Uplift-net for simplicity.
The sum of $Logits_{ctrl}$ and $V_{uplift}$ refers to the purchase logits when making an additional request, \ie, $Logits_{trt}$.
Finally, we obtain the purchase rate estimation of the CONTROL group and the TREATMENT group $P_{ctrl}$, $P_{trt}$, and the estimation of $CATE_P$: 
\vspace{-0.5em}
\begin{gather}
P_{ctrl} = Sigmoid(Logits_{ctrl}), \label{eq:pred0} \\
P_{trt} = Sigmoid(Logits_{ctrl} + V_{uplift}), \label{eq:pred1} \\
\widehat{CATE_P} = P_{trt} - P_{ctrl}, \label{eq:pred2} 
\end{gather}
where $\widehat{CATE_P}$ is the estimation of purcharse uplift $CATE_P$. Since $V_{uplift}$ is positively correlated with $\widehat{CATE_P}$, we substitute $V_{uplift}$ for  $\widehat{CATE_P}$ as the basis for request decision in the \textbf{DRP} module. 

\vpara{Loss.} 
We use the purchase behavior as labels for users in the CONTROL group and the TREATMENT group, \ie, $Label_{ctrl}$ and $Label_{trt}$, which is equal to one if the purchase behavior happens otherwise zero. We use \textit{cross-entropy} as the loss function.  
\begin{gather}
Loss_{ctrl} = -Label_{ctrl}\log(P_{ctrl}) - (1-Label_{ctrl})\log(1-P_{ctrl}), \notag \\
Loss_{trt} = -Label_{trt}\log(P_{trt}) - (1-Label_{trt})\log(1-P_{trt}), \\
Loss_{all} = Loss_{ctrl} + Loss_{trt}. \notag \label{eq:loss}
\end{gather}

\subsubsection{\textbf{Dynamic Request Programming (DRP)}} \label{sec:DRP}

Due to the constraints on computational resource and communication bandwidth in industrial systems, 
one needs to trade off the rewards and the resource consumption brought by requests. 
In industrial scenarios, it is common to set an upper bound $\theta$ on the resource consumption of additional requests over a period of time. 
In this context, \textbf{DRP} aims to maximize the purchase uplift while keeping the resource consumption of additional requests under the upper bound $\theta$. The problem can be formulated as follows:
\begin{gather}
\text{Maximize} \sum_{X_i \in \mathcal{X}} z_i CATE_P(X_i), \label{eq:knapsack0} \\
\text{subject to} \sum_{X_i \in \mathcal{X}} z_i CATE_Q(X_i) < \theta,  \label{eq:knapsack1} \\
CATE_Q(X_i) = E[Q_i(1) \vert X_i]-E[Q_i(0) \vert X_i]=\lambda, \label{eq:knapsack2}
\end{gather}
where $CATE_P(X_i)$ and $z_i$ has been defined in \textit{Eq.} \ref{eq:cate0}-\ref{eq:cate1}, $Q_i(1)$, $Q_i(0)$ denote the resource consumption when inserting a  request or not, respectively, and $CATE_Q(X_i)$ refers to the uplift of resource consumption brought by an additional request in the case $X_i$. Since each request consumes almost the same, $CATE_Q(X_i)$ can be approtimated by a constant $\lambda$.
This is a typical binary knapsack problem since a request either occurs or not. Furthermore, since the resource consumption of each request is almost constant, we can use the commonly adopted greedy algorithm to obtain an optimal solution.
\begin{gather}
g(X_i) = \frac{CATE_P(X_i)}{CATE_Q(X_i)} \approx \frac{1}{\lambda} \widehat{CATE_P}(X_i). \label{eq:knapsack}
\end{gather}
A straightforward solution with the greedy algorithm for request programming is to use $g(X_i)$, which is the ratio of purchase uplift to resources consumption as in \textit{Eq.} \ref{eq:knapsack}, as the basis to rank all the cases where request decisions need to be made in a period of time, and select the top $\mathit{M}\%$ cases. Since  $V_{uplift}$ is positively correlated with the purchase uplift estimation $\widehat{CATE_P}$, we can substitute $V_{uplift}$ for $g(X_i)$ as the ranking score. However, we cannot obtain the $V_{uplift}$ set of all the cases until the end of a time period, while we need to timely make a decision for each case. As an approximation, we take the $V_{uplift}$ scores of all the cases in the previous time period as guidance. Specifically, if we aim to pick up the top $\mathit{M}\%$ requests in the current period, we set the threshold as the minimum score of the top $\mathit{M}\%$ cases in the previous time period as an approximation.

\begin{table*}[h]
\centering
    \caption{ Offline experiment on a large-scale industrial dataset. 
    We conduct two-sided test and p-value $< 0.05$ shows the improvement of AdaRequest over the strongest baseline (underlined) is statistically significant.}
{\setlength{\tabcolsep}{0.6em}\renewcommand{\arraystretch}{0.9}\begin{tabular}{l ccccccccc}
\toprule

  Metric & AdaRequest & OneModel & TwoModel  & ClassTrans\footnotemark &  Greedy & RandR & PoolR  & StaticR  & p-value   \\
    \midrule
Qini AUUC  $\uparrow$ &	$\mathbf{1.8288}$ &  $\underline{1.4767}$ &   $ 1.4007 $   &	$1.3558$ &	$0.4802$
 &	$0.0398$ &	$-3.0404$ &	$-1.6724$ &	$2.3130 \times 10^{-6}$ \\
 
 Qini (50) $\uparrow$
 &	$\mathbf{4.3450}$ &  $\underline{3.8859}$ &   $ 3.8061 $   &	$3.7662$ &	$2.8371$ &	$1.9209$
    &	$-2.3353$ &	$0.1229$ &	$3.0514 \times 10^{-3}$ \\
 
  MSE Y* $\downarrow$\footnotemark &	$\mathbf{4.3497}$    & $4.3526$   
 &	$\underline{4.3524}$ &	- &	- &	- &	-
 &	- &	$7.5479 \times 10^{-6}$ \\
 
 \midrule
 
 AUC $\uparrow$ &	$\mathbf{0.8145}$ &	$0.7980$ &  $\underline{0.7980}$    
 &	- &	$0.7974$ &	$0.5017$ &	$0.4706$ &	$0.4817$
    &	$3.5235 \times 10^{-9}$ \\
 
 MSE $\downarrow$ &	$\mathbf{1.0833}$ &  $1.0843$ & $\underline{1.0839}$ &	-
 &	$1.0840$ &	$504.9$ &	$711.9$ &	$846.6$ &	$6.3846 \times 10^{-3}$ \\

    \bottomrule
\end{tabular}}
    \label{tab:comparison}
\end{table*}
%

\section{Experiments}

We conduct extensive offline and online experiments to answer the following research questions:
\begin{itemize}[leftmargin=*]
    \item \textbf{RQ1}: How does the model of AdaRequest perform as compared to baseline methods?
    \item \textbf{RQ2}: How do different model building blocks and critical behavior features affect the effectiveness of AdaRequest?
    \item \textbf{RQ3}: What is AdaRequest's online performance when deployed in Mobile Taobao?
\end{itemize}

\subsection{Experimental Setup} \label{sec:expset}


\vpara{Dataset.} 
For offline experiments, we collected a large-scale industrial dataset \textit{TaoBao Request} from the homepage waterfall RS \textit{Guess You Like} in Mobile Taobao, containing $1.76\times10^8$ user behavior and request records.
More details can be found in Appendix \ref{sec:data}.

\vpara{Baselines.}
To the best of our knowledge, there are no existing methods doing precisely the same task as ours. Therefore, we construct model-based methods based on representative related works and incorporate non-adaptive request strategy as baselines.
\begin{itemize}[leftmargin=*]
    \item \textbf{StaticR, PoolR, RandR}. StaticR and RandR inserts a fixed number of requests between adjacent page breaks in a uniform and random manner, respectively. PoolR inserts requests based on the number of candidate items on edge.
    \item \textbf{Greedy}. Greedy uses all features as in AdaRequest and uses average pooling plus DNNs for prediction. Greedy directly pursues the absolute purchase rate and uses the purchase rate rather than the uplift as the criterion for request decision.
    \item \textbf{ClassTrans~\cite{zhang2021unified}}. ClassTrans differs from Greedy by predicting the uplift directly estimated by the class transformation method.
    \item \textbf{TwoModel~\cite{nie2021quasi}}. Different from ClassTrans, TwoModel incorporates two independent models trained on the control group and the treatment group separately. The uplift is estimated as the difference of two models.
    \item \textbf{OneModel~\cite{zhao2017uplift}}. OneModel differs TwoModel by taking a $0/1$ signal indicating whether there is an additional request imposed currently as input. The uplift is estimated as the prediction difference of two forward passes with different signal values.
\end{itemize}

\vpara{Evaluation Metrics.}
\begin{itemize}[leftmargin=*]
    \item 
    \textit{Offline Evaluation.} 
    We consider two kinds of metrics for offline evaluation, \ie, uplift related (Qini AUUC, Qini 50, and MSE $Y^*$) and purchase rate related (AUC and MSE). For simplicity, throughout the experiment section, we use the term \underline{\textbf{uplift}} to denote the difference of predicted purchase rates if a request is made and if not. Uplift measures the real purchase rate gain rather than the absolute purchase rate, as illustrated in \textit{Eq.} \ref{eq:cate1}. The details of metrics can be found in Appendix \ref{sec:app_metrics}.
Since non-adaptive strategies RandR, PoolR, and StaticR can only give binary outputs $1/0$ when they decide to or not to insert a request, we take these binary outputs directly as their predictions for uplift or purchase rate when computing above metrics.

    \item \textit{Online Evaluation.} We use two online metrics for evaluation, \ie, purchase rate in the subsequent $N$ items (PR in $N$), and Gross Merchandise Volume (GMV). We measure the GMV \wrt different Query-Per-Second constraints, and obtain the GMV-QPS plot, which reveals the platform-level revenue of models under different resource constraints.
\end{itemize}
For simplicity, we omit the scales of metric values, which are $10^{-4}$ for Qini AUUC, $10^{-4}$ for Qini 50, $10^{-3}$ for MSE $Y^*$, $10^{-3}$ for MSE, and $10^{-3}$ for PR in $N$.


\begin{figure}[!t] \begin{center}
	\includegraphics[width=0.98\linewidth]{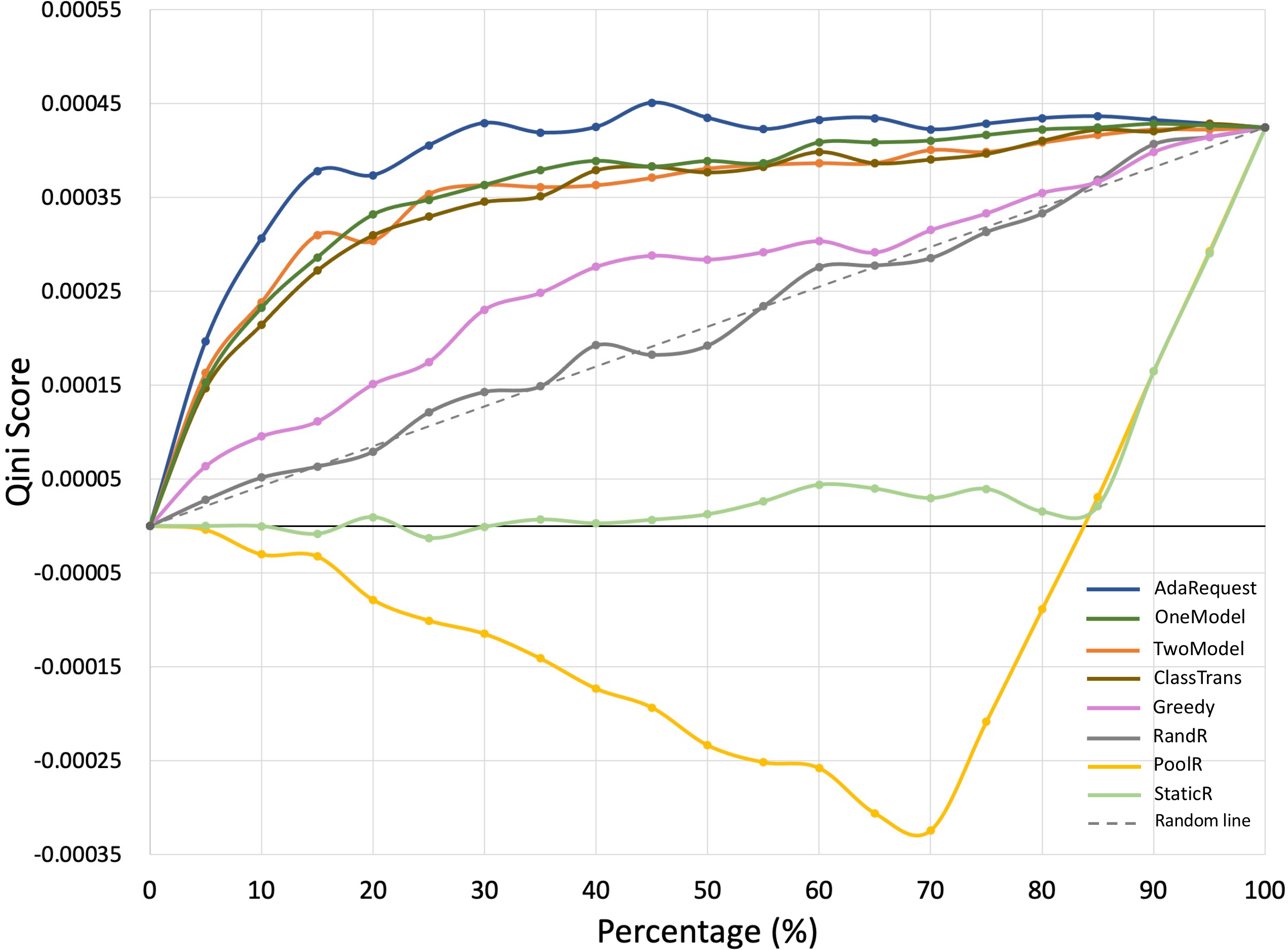}
    \caption{
    Qini curves which reveal the uplift prediction performance \wrt different request resource constraints.
    	}
    \label{fig:gini_offline}
\vspace{-0.5cm}
\end{center} \end{figure}

\subsection{Overall Performance (RQ1)} \label{sec:overper}

Table \ref{tab:comparison} shows the prediction performance of diverse baselines and AdaRequest.
Not surprisingly, the Greedy method that considers user behavior and context outperforms simple non-adaptive baselines (PoolR, StaticR, and RandR). 
However, Greedy solely pursues the absolute purchase rate brought by a request, resulting in that its performance on uplift prediction (\eg, Qini AUUC) is one order of magnitude worse than baselines (ClassTrans, TwoModel) that explicitly purses the uplift. 
ClassTrans directly predicts the uplift in one step under the Conditional Independent Assumption $(CIA)$ between the treatment and user attribute/behaviors. 
Obviously, this assumption might not hold in real-world systems due to the ubiquitous data biases. 
For example, with more interactions in RS, active users inevitably obtain more treatments and dominate the treatment group data.
In this case, treatment is no longer independent of user behavior in the observational training dataset.
Therefore, although ClassTrans can beat Greedy, it is still inferior compared to TwoModel, which is less affected by $CIA$ by using two independent models to estimate the purchase rate of the treatment group and the control group, respectively. 
OneModel achieves comparable performance with TwoModel and outperforms it on Qini related metrics. We attribute the improvement to the shared parameters for the treatment group and the control group, leading to consistent and less-overfitting training with more data samples. 

\begin{sloppypar}

\textbf{AdaRequest} consistently yields the best performance across different metrics. 
Remarkably, AdaRequest improves the best-performing baseline by relatively 29.5\% in terms of Qini AUUC. 
We attribute the performance gain to the superiority of the user intention capturing module (CUBE), and the counterfactual request reward estimator module (CREST). 
Specifically, CUBE perform user real-time intention mining by comprehensively and chronologically modeling users' diverse behavior and interaction commodity sequences. Based on attention-based neural networks and using candidate items as the query, CUBE comparing user's real-time intention with their historical interest.
Furthermore, different from TwoModel which independently models the treatment and the control groups using two models, CREST reuses the backbone CUBE for the representation learning of all users, leading to robust and consistent user modeling.

\end{sloppypar}

\vpara{Uplift \wrt Resource Limitation.} 
We plot the Qini curves of different models in Figure \ref{fig:gini_offline}.
The proposed AdaRequest model shows consistent quantitative improvement over all baselines under a variety of resource limitation configurations, which demonstrates the rationality of our analysis and the merits of our model.
By taking an in-depth analysis of different resource thresholds, we found that the performance gain of AdaRequest over the best-performing baseline is significantly larger when the resource is more limited, especially when the percentage $K<50$. In the deployed system, the empirical maximum resource limitation nearly corresponds to $K=50$. These results further reveal the practical value of AdaRequest on real-world resource-limited scenarios.

\begin{table}[!t]
\centering
    \caption{ Ablation Study on Model Architecture. }
{\setlength{\tabcolsep}{0.3em}\renewcommand{\arraystretch}{0.9}\begin{tabular}{l ccc}
\toprule

  Model & Qini AUUC $\uparrow$ &  Qini (50) $\uparrow$  &  MSE  $Y^*$ $\downarrow$   \\
    \midrule
AdaRequest  &	$\mathbf{1.8288}$  &	$\mathbf{4.3450}$  &	$\mathbf{4.3497}$ \\
w/o CUBE  &	$1.5894$  &	$4.0260$  &	$4.3504$  \\
w/o BShare  &	$1.6629$  &	$4.0852$  &	$4.3521$ \\
w/o CREST w. CT  &	$1.4722$  &	$3.9057$  &	- \\
w/o CREST w. CO  & $1.7300$  &	$4.2053$  &	$4.3522$ \\
w/o Uplift  &	$0.5339$  &	$3.3332$  &	- \\
    \bottomrule
\end{tabular}}
    \label{tab:modelablation}
\vspace{-0.2cm}
\end{table}
%

\subsection{Ablation Studies (RQ2)}

\textbf{Ablating Architectures.}  To show the effectiveness of different components in AdaRequest, we remove one component at a time and obtain multiple architectures. The results are shown in Table \ref{tab:modelablation}. We use w/o and w. to denote without and with, respectively.
\begin{itemize}[leftmargin=*]
    \item \textbf{w/o CUBE} means that we aggregate all features using mean-pooling rather than the proposed CUBE. The large-margin performance gap reveals that comprehensive and in-depth user understanding in CUBE is a critical contributing factor to request strategy design.
    \item \textbf{w/o BShare}. This means we imitate the structure of TwoModel by using two CUBEs to model user behavior separately, and use the outputs of the two CUBEs as inputs to the Control-net and the Uplift-net in CREST respectively. The performance drop demonstrates the effectiveness of the design of backbone sharing which lead to consistent and less-overfitting training.
    \item \textbf{w/o CREST, w. CT}. We substitute CREST with the method of uplift prediction in ClassTrans\cite{zhang2021unified}. The poorer performance shows the merits of CREST in uplift prediction.
    \item \textbf{w/o CREST, w. CO}. This means that we substitute CREST with the method of uplift prediction in OneModel. 
    We take whether we insert a request as the $0/1$ condition signal and take the prediction difference between the two forward passes inputing different condition signals as the uplift. 
    The poorer performance show the superiority of CREST as well.
    \item \textbf{w/o Uplift}. This means we solely pursue the absolute purchase rate without any consideration of the uplift. Analogous to the findings in Section \ref{sec:overper}, this greedy strategy leads to a significant performance drop.
\end{itemize}

\addtocounter{footnote}{-1}\footnotetext{ClassTrans does not predict the purchase rate and thus without AUC and MSE.}
\addtocounter{footnote}{+1}\footnotetext{MSE Y* is only valid for methods that predict uplift. MSE Y* of ClassTrans is not meaningful due to its distinctive uplifit predition distribution.}

\begin{table}[!t]
\centering
    \caption{ Ablation Study on Feature System.}
{\setlength{\tabcolsep}{0.5em}\renewcommand{\arraystretch}{0.9}\begin{tabular}{l cccc}
\toprule

  Model & Qini AUUC $\uparrow$ &  Qini (50) $\uparrow$ &  AUC $\uparrow$  &  MSE $\downarrow$   \\
    \midrule
AdaRequest  &	$\mathbf{1.8288}$  &	$\mathbf{4.3450}$  &	$\mathbf{0.8145}$  &	$\mathbf{1.0833}$ \\
w/o $F_{context}$ &	$1.7480$  &	$4.3251$  &	$0.8107$  &	$1.0843$ \\
w/o $F_{exp}$  &	$1.6980$  &	$4.1253$  &	$0.8110$  &	$1.0833$ \\
w/o $F_{clk}$  &	$1.7899$  &	$4.3250$  &	$0.8134$  &	$1.0834$ \\
w/o $F_{sclk}$  &	$1.8188$  &	$4.2852$  &	$0.8137$  &	$1.0833$ \\

    \bottomrule
\end{tabular}}
    \label{tab:featablation}
\vspace{-0.2cm}
\end{table}
%

\vpara{Ablating Feature Schema.} To reveal how the features in the proposed feature schema contribute to AdaRequest, we remove one kind of feature at one time. Table \ref{tab:featablation} lists the results. In a nutshell, removing any feature leads to a performance drop which reveals the effectiveness of the proposed feature schema.
\begin{itemize}[leftmargin=*]
    \item \textbf{w/o} $F_{context}$. This means that we disregard coarse-grained context features such as the view depth and the number of clicks. Apparently, these features are essential for analyzing whether the user is interested in the current recommendations.
    Therefore, these features largely contribute to the performance.
    \item \textbf{w/o} $F_{exp}$. $F_{exp}$ indicates the exposed item sequence in current session, including items users are interested in and, perhaps more importantly, items users are not interested in. Along with user click sequences, we can effectively measure whether the candidate items still satisfy the user's intentions. As such, $F_{exp}$ contributes the most Qini AUUC and Qini (50) scores.
    \item \textbf{w/o} $F_{clk}$. $F_{clk}$ denotes the long-term clicked item sequence. Due to the sparsity of user behaviors, it might be error-prone to capture users' inherent interests in the current session. Therefore, long-term behaviors lay the foundation of analyzing the change of users' intentions. The performance drop after removing the long-term behaviors is evidence of this analysis.
    \item \textbf{w/o} $F_{sclk}$. $F_{sclk}$ includes behavior sequence in the current session, which is more real-time and can be essential to analyze users' current intention change. Therefore, although the behavior data within the scope of a session can be sparse and sometimes noisy, it still has a moderate contribution to the Qini (50) metric. 
\end{itemize}

\begin{table}[!t]
\centering
    \caption{ Direct benefit from inseted requests of AdaRequest and online baselines. PR in $N$ indicates the purchase rate in the following $N$ items. }
{\setlength{\tabcolsep}{0.75em}\renewcommand{\arraystretch}{0.9}\begin{tabular}{c cccccc}
\toprule

  PR in $N$  & NoR  &  RandR &  PoolR  &  StaticR & AdaRequest  \\
    \midrule
10  &	0.889  &	$1.174$  &	\underline{1.177}  &	$1.130$  &	$\mathbf{2.289}$   \\
20  &	1.660  &	\underline{2.197}  &	$2.164$  &	$2.045$  &	$\mathbf{4.173}$  \\

    \bottomrule
\end{tabular}}
    \label{tab:buy_online}
\vspace{-0.2cm}
\end{table}
%

\subsection{Online Performance and Analysis (RQ3)}

Currently, the AdaRequest has been fully deployed on the homepage waterfall RS \textit{Guess You Like} in Mobile Taobao, and serves billions of users. 
We quantitatively show the benefit of deployed AdaRequest as follows. 
The online experiment lasted 7 days and covered more than 10 million users.
The qualitative results (\textbf{Case Study}) can be found in Appendix \ref{sec:casestudy}.

\vpara{Deployment.} 
We deploy the AdaRequest in the waterfall RS \textit{Guess You Like} in Mobile Taobao on edge. 
The AdaRequest takes about $500ms$ for each run, which is a very efficient operation.
We use the following four types of user behaviors as the trigger of AdaRequest: 1). user clicks on an item. 2). user slides down more than $n$ items, $n=6$. 3). user slides down and stops. 4). user deletes an item. 
Note that the model in AdaRequest is deployed almost entirely on edge, except for the item embedding matrixes, brand embedding matrixes, and category embedding matrixes inside the model.
Since the parameter size of these three embedding matrixes is too large, we deploy them on the cloud. When the cloud has determined the new recommended items in response to a request, it would retrieve the embedding related to the items in these three embedding matrixes, send them back to the edge along with the items. These embeddings would be used as input directly when the model on edge is invoked.

\vpara{Direct Benefit.} 
The direct benefit can be reflected by the average purchase rate in the following $N$ items after a request is imposed, \ie, \textbf{PR in $N$}. Previous online baselines include RandR, PoolR, StaticR, and No Request (\textbf{NoR}) which means sole paging request mechanism without any additional requests, as illustrated in Section \ref{sec:expset}.
The results are listed in Table \ref{tab:buy_online}. Due to confidentiality considerations, we solely show the results with $N \in {10, 20}$. We observe that the direct online benefit brought by AdaRequest is consistently better than those of previous online baselines. 
The results are consistent with the findings in offline experiments, and, more importantly, demonstrate the practical value of AdaRequest in real-world scenarios.

\begin{figure}[!t] \begin{center}
	\includegraphics[width=1.\linewidth]{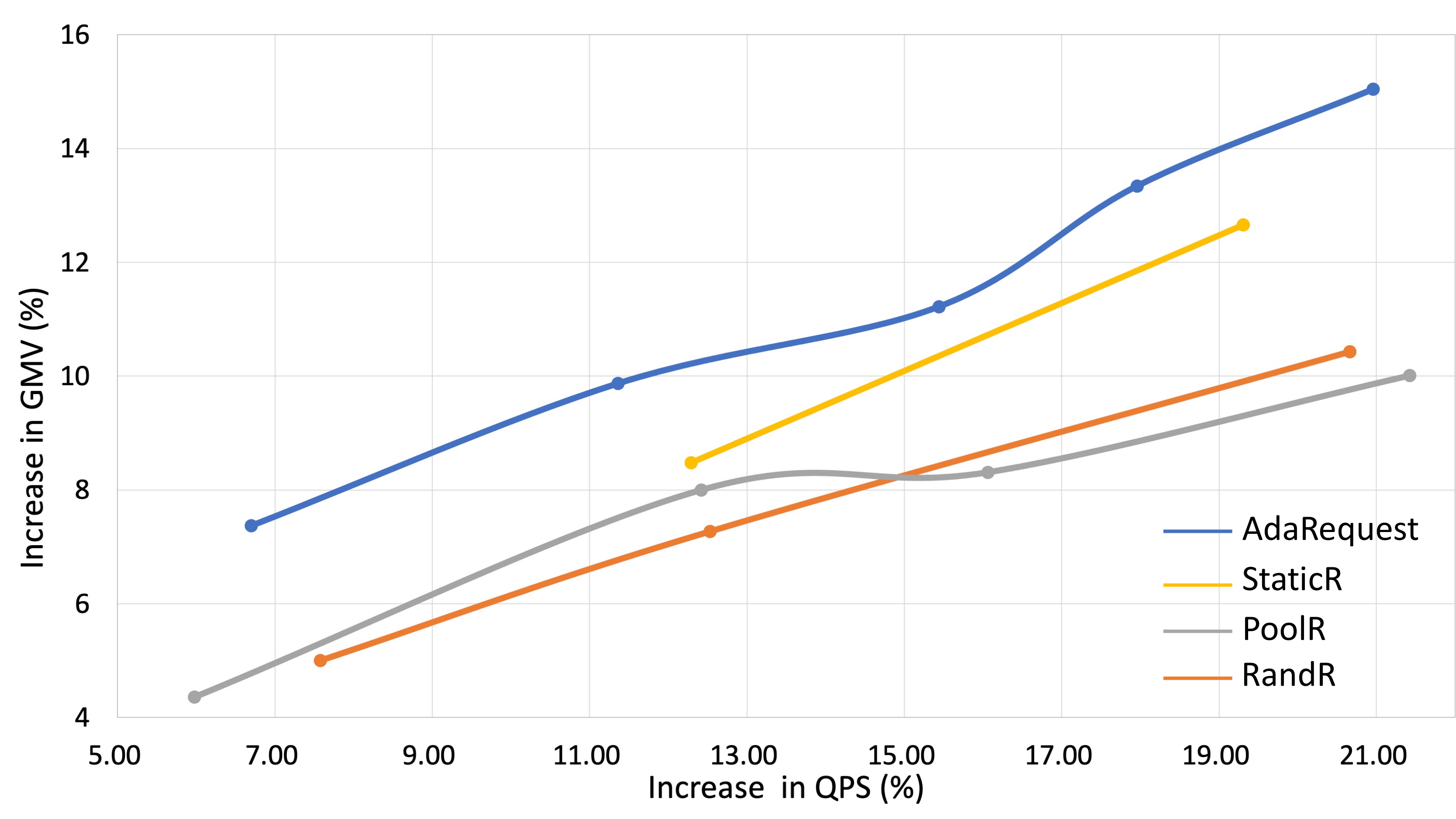}
    \caption{
    Online A/B test which reveals GMV improvement \wrt increased Query-Per-Second (QPS) upper bound.
    	}
    \label{fig:gmv_online}
\vspace{-0.4cm}
\end{center} \end{figure}

\vpara{Platform Benefit.} The direct benefit reflects the timely benefit brought by additional requests. It is noteworthy that in some situations, the direct benefit might sacrifice the benefit of other components in the whole platform and even hurt the overall benefit in the long run. Therefore, we take a step further to reveal the overall platform benefit through online A/B test. Remarkably, 
the platform's GMV improvement of AdaRequest reaches over $\mathbf{3\%}$ compared to the previous online baseline StaticR. 


To reveal the online platform benefit \wrt different QPS constraints, we view the QPS of No Request(NoR) as the starting point, and vary the relative QPS increase from $5\%$ to $21\%$. Through A/B test, we obtain the GMV improvement of AdaRequest over the No Request baseline for each QPS constraint. We also obtain the results of three online baselines, including RandR, PoolR and StaticR, for comparison. It is noteworthy that AdaRequest and other baselines work with EdgeRec~\cite{Gong_Jiang_Feng_Hu_Zhao_Liu_Ou_2020} online.
Note that since the online QPS cannot be precisely controlled, we cannot compare the effects of different methods in the same abscissa.
The results are shown in Figure \ref{fig:gmv_online}. We can see that AdaRequest achieves consistent GMV improvement over other baselines across different QPS constraints. These results show that AdaRequest not only improves the timely benefit reflected in users' short-term purchase rate, but also contributes to the overall platform revenue in the long run.

\section{Conclusion} \label{sec:conclusion}

In this paper, we investigate a novel research and industrial problem, \ie, Intelligent Request Strategy Design, aiming to capture users' real-time intention change, and accordingly request the server for candidate item update. We summarize the critical challenges, including the non-trivial user intention mining from real-time implicit behavior features, accurately measuring the request reward, and request programming \wrt server Query-Per-Second (QPS) constraint. We propose a novel framework AdaRequest to address these challenges. 
Extensive offline and online experiments demonstrate the significance of this problem, the rationality of our analysis, and the effectiveness of AdaRequest.

\section{ACKNOWLEDGMENTS}

\begin{sloppypar}
This research is supported by the Starry Night Science Fund of Zhejiang University Shanghai Institute for Advanced Study (SN-ZJU-SIAS-0010).
\end{sloppypar}

\balance

\bibliographystyle{ACM-Reference-Format}
\bibliography{sections/9.citations}

\appendix

\begin{table}[h]
\caption{Offline Dataset Statistics.}
\centering
\begin{tabular}{ccc}
\toprule
\#Num of Records & \#Num of Users & \#Num of Items   \\
    \midrule
    $1.76\times10^8$  & $1.01\times10^7$ & $2.98\times10^7$ \\
    \bottomrule
\end{tabular}%
    \label{tab:staData}
\end{table}
%

\section{Appendix}
\label{sec:appendix}

\subsection{Dataset} \label{sec:data}
The order of magnitude of the records in the dataset Taobao Request reach $10^8$ and the detailed statistics in listed in Table \ref{tab:staData}. Specifically, it includes information associated with candicate items and the items that have been exposed, as well as users' browsing history, click history, and coarse-grained context features (\eg, view depth), and the number of clicks. In addition, the dataset includes user purchase records in the current session as the training labels.

Note that in a particular time point, we cannot silmutaneously observe users' behaviors when there is a request made or not made. 
Therefore, in order to train models, we divide users into the CONTROL group, where there is no request made, and the TREATMENT group, where requests are randomly inserted, in the data collection phase. We keep the number of data samples in the control group and the treatment group nearly the same to prevent model training bias caused by data imbalance. We split data samples into training and testing at the ratio of 85\% and 15\%, respectively.

\subsection{Metrics}
\label{sec:app_metrics}

\begin{itemize}[leftmargin=*]
    \item 
    \textit{Offline Evaluation.} 
        \begin{itemize}[leftmargin=*]
        \item \textbf{Qini (50).} It is infeasible to measure the uplift for a particular data sample since we solely have the factual purchase rate when a request is made or not made. Therefore, we measure the statistical dispersion of the predicted uplift scores for the samples in the control group and the treatment group.
        Specifically, we adopt Qini score \cite{radcliffe2007using}: \begin{align}
            Q(\phi)=\frac{n_{t, y=1}(\phi)}{N_{t}}-\frac{n_{c, y=1}(\phi)}{N_{c}}
        \end{align}
        where we measure the Top $\phi \%$ samples with the highest predicted uplift scores. It is noteworthy that $\phi$ corresponds precisely to the request resource constraint as defined in Section \ref{sec:DRP}. Therefore, Qini score reflects the performance in uplift prediction and selecting cases with high ratio $g(X_i)$ under a particular resource limitation. We adopt $\phi = 50$, which is used in industrial scenarios. $n_{t, y=1}$, $n_{c, y=1}$ refer to the number of cases where purchases eventually happened in the treatment and the control group of the Top $\phi \%$ samples, respectively. $N_{t}$ and $N_{c}$ refer to the total number of cases in the treatment and the control group, respectively.
        \item \textbf{Qini AUUC.} Different from Qini 50, we vary $\phi$ and obtain different Qini scores, resulting in a Qini curve. We measure the area between the Qini curve of a particular model and the theoretical Qini curve of random request strategy. Qini AUUC reflects an overall performance covering different request resource constraints.
        \item \textbf{MSE $Y^*$.} We borrow the MSE Metric Based on Y* from \cite{gutierrez2017causal} to measure the uplift.
        \item \textbf{AUC \& MSE.} Two widely known metrics to measure the purchase rate prediction performance.
    \end{itemize}
    \item 
    \textit{Online Evaluation.} 
    \begin{itemize}[leftmargin=*]
        \item \textbf{PR in $N$} measure the average purchase rate in the next $N$ items after an additional request being inserted, which reflects the timely reward brought by a request. 
        \item \textbf{GMV} is the total amount of online income during a period. 
    \end{itemize}
\end{itemize}

\subsection{Feature Details} \label{sec:featdetail}

In the deployed system, we consider heterogeneous features for a comprehensive understanding of user intention and whether the candidate items satisfy the current user. As illustrated in Section \ref{sec:overview}, all features can be roughly divided into two kinds, \ie, fine-grained user behavior features and coarse-grained context features. We listed the detailed features of these two kinds in Table \ref{tab:featbehavior} and Table \ref{tab:featcontext}, respectively.
\begin{itemize}[leftmargin=*]
    \item \textbf{Behavior Feature}. We consider three kinds of features, \ie, items exposed in the current session, items clicked in the current session, items historically clicked. For each item, we consider heterogenous categorical features and continuous features for item representation learning, which lays the foundation of futher intention understanding. For example, item category could help the model identify whether the user becomes interested in a different category of items from clicked and exposed item sequences.
    \item \textbf{Context Feature}. Besides behavior features, we also take heterogeneous contenxt features, including user features, candidate item features, and session features, into consideration. Candidate item representation learning is similar to that of interacted items, \ie, with similar feature inputs. Differently, we consider the position of a candidate item in the candidate pool as input. User features include inherent ones such as gender/age, and statistical ones such as click-through-rate, and estimated ones such as purchasing power. Session features mainly include statistical ones related to page and request.
\end{itemize}

\begin{table}[h]
\centering
    \caption{ Fine-grained user behavior features. }
{\setlength{\tabcolsep}{0.15em}\renewcommand{\arraystretch}{0.75}\begin{tabular}{ll lc}
\toprule
\footnotesize
Item Type &  Feat. Name  & Description  &  Feat. Type  \\
    \midrule
\multirow{7}{*}{Exposed}
  &	exp\_item\_fea\_seq  &	 CTR, CVR, etc.   &	embedding  \\
  &	exp\_item\_id\_seq  &	 ID  &	bucketize  \\
  &	exp\_item\_cat\_seq  &	 Category  &	bucketize  \\
  &	exp\_item\_brand\_seq  &	 Brand  &	bucketize  \\
  &	exp\_item\_pos\_seq  &	 Position in Page  &	bucketize  \\
  &	exp\_item\_page\_seq  &	 Page Num.  &	bucketize  \\
  &	exp\_item\_price\_seq  &	 Price Level  &	bucketize  \\
    \midrule
\multirow{7}{*}{\shortstack[l]{Clicked \\ in History}}
  &	clk\_item\_fea\_seq  &	 CTR, CVR, etc.  &	embedding  \\
  &	clk\_item\_id\_seq  &	 ID  &	bucketize  \\
  &	clk\_item\_cat\_seq  &	 Category  &	bucketize  \\
  &	clk\_item\_brand\_seq  &	 Brand  &	bucketize  \\
  &	clk\_item\_pos\_seq  &	 Position in Page  &	bucketize  \\
  &	clk\_item\_page\_seq  &	 Page Num.  &	bucketize  \\
  &	clk\_item\_price\_seq  &	 Price Level  &	bucketize  \\
\midrule
\multirow{7}{*}{\shortstack[l]{Clicked \\ in Session}}
  &	sclk\_item\_fea\_seq  &	 CTR, CVR, etc.  &	embedding  \\
  &	sclk\_item\_id\_seq  &	 ID  &	bucketize  \\
  &	sclk\_item\_cat\_seq  &	 Category  &	bucketize  \\
  &	sclk\_item\_brand\_seq  &	 Brand  &	bucketize  \\
  &	sclk\_item\_pos\_seq  &	 Position in Page  &	bucketize  \\
  &	sclk\_item\_page\_seq  &	 Page Num.  &	bucketize  \\
  &	sclk\_item\_price\_seq  &	 Price Level  &	bucketize  \\

    \bottomrule
\end{tabular}}
    \label{tab:featbehavior}
\end{table}
%

\begin{table}[h]
\centering
    \caption{ Coarse-grained context features. }
{\setlength{\tabcolsep}{0.15em}\renewcommand{\arraystretch}{0.65}\begin{tabular}{l lc}
\toprule

Feat. Name  & Description  &  Type  \\
    \midrule
  cand\_item\_fea\_seq  &	 CTR, CVR, etc.  &	embedding  \\
  cand\_item\_id\_seq  &	 ID  &	bucketize  \\
  cand\_item\_cat\_seq  &	 Category  &	bucketize  \\
  cand\_item\_pos\_seq  &	Position in Pool  &	bucketize  \\
  cand\_item\_page\_seq  &	 Page &	bucketize  \\
  cand\_item\_brand\_seq  &	 Brand  &	bucketize  \\
  cand\_item\_price\_seq  &	 Price Level  &	bucketize  \\
  \midrule
user\_os  &	Operating System  &	bucketize  \\
user\_gender  &	Gender  &	bucketize  \\
user\_age\_level  &	Age Level  &	bucketize  \\
user\_purchase\_level  &	Purchasing Power  &	bucketize  \\
avg\_pv  &	Page Review Num.  &	bucketize  \\
avg\_clk  &	Click Num.  &	bucketize  \\
avg\_pay  &	Purchase Num.  &	bucketize  \\
avg\_ctr  &	Click-through-rate  &	bucketize  \\
avg\_cvr  &	Conversion Rate  &	bucketize  \\
\midrule
user\_hour  &	Curent Time  &	bucketize  \\
visit\_depth  &	Browsing Depth  &	bucketize  \\
page\_flow\_pos\_level  &	Position in Page  &	bucketize  \\
buffer\_left\_item\_level  &	Item Pool Size  &	bucketize  \\
sessionpv\_happened\_req  &	In-session Request Num.  &	bucketize  \\
cur\_page\_have\_req  &	In-page Request Num.  &	bucketize  \\
session\_last\_req\_dist  &	Dist. to the Last Request  &	bucketize  \\
session\_last\_req\_delay  &	Time to the Last Request  &	bucketize  \\
last\_session\_ipv\_interval  &	Time to the Last Click  &	bucketize  \\
last\_session\_ipv\_dist  &	Dist. to the Last Request  &	bucketize  \\
session\_ipvs\_len  &	In-session Click Num.  &	bucketize  \\

    \bottomrule
\end{tabular}}
    \label{tab:featcontext}
\end{table}
%

\subsection{Case Study} \label{sec:casestudy} 
\begin{figure}[h] \begin{center}
	\includegraphics[width=1.\linewidth]{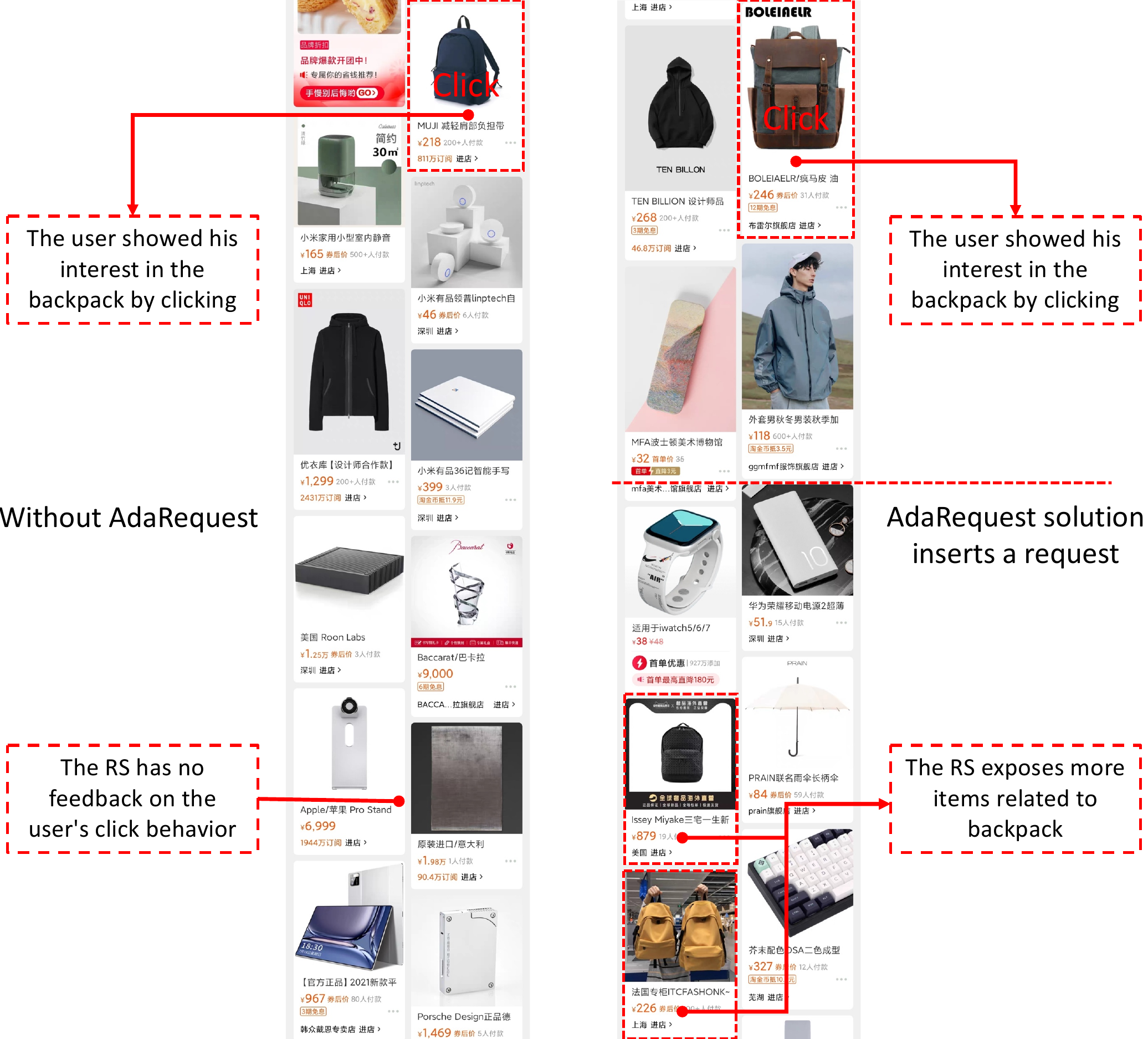}
    \caption{
    Case study by visualizing two cases from Mobile Taobao with (right) and without (left) AdaRequest deployed.
    	}
    \label{fig:case}
\end{center} \end{figure}

To vividly depict the benefit of AdaRequest in deployed systems, we conduct a case study on Mobile TaoBao, as shown in Figure \ref{fig:case}. Without AdaRequest, the system has no capability to give immediate feedback to users' real-time behaviors and the underlying intentions. Therefore, when a user clicks a backpack, potentially showing his interest in related items, the candidate items might fail to satisfy the user's need. Therefore, the user might scroll through a lot of items until he/she meets a physical page break, or just leave the platform in the middle feeling disinterested, hurting users' experiences. On the contrary, as shown in Figure \ref{fig:case}, AdaRequest successfully captures users' intention change after he/she clicks a backpack, and sends a request to update the candidate items. After the request, the system exposes related items (\ie, backpacks) in order to better satisfy users' new intentions.

\end{document}